\documentclass{article}
\usepackage{kr}

\usepackage{times}
\usepackage{soul}
\usepackage{url}
\usepackage[utf8]{inputenc}
\usepackage[small]{caption}
\usepackage{graphicx}
\usepackage{amsmath}
\usepackage{amsthm}
\usepackage{booktabs}
\usepackage{colortbl}
\usepackage{algorithm}
\usepackage{algorithmic}
\usepackage{xspace}
\usepackage[obeyFinal]{todonotes}

\graphicspath{{./image/}}
\newtheorem{theorem}{Theorem}
\newtheorem{example}[theorem]{Example}
\newtheorem{definition}[theorem]{Definition}
\newtheorem{proposition}[theorem]{Proposition}
\newtheorem{problem}[theorem]{Problem}

\usepackage{amssymb,amsfonts}
\usepackage{tikz}
\usetikzlibrary{positioning,shapes,arrows,fit}
\usepackage{standalone}

\newcommand{\Dmc}{\ensuremath{\mathcal{D}}\xspace}
\newcommand{\Imc}{\ensuremath{\mathcal{I}}\xspace}
\newcommand{\Kmc}{\ensuremath{\mathcal{K}}\xspace}
\newcommand{\Pmc}{\ensuremath{\mathcal{P}}\xspace}
\newcommand{\Tmc}{\ensuremath{\mathcal{T}}\xspace}
\newcommand{\Vmc}{\ensuremath{\mathcal{V}}\xspace}
\newcommand{\Wmc}{\ensuremath{\mathcal{W}}\xspace}

\newcommand{\Imf}{\ensuremath{\mathfrak{I}}\xspace}

\newcommand{\exa}{\ensuremath{\mathsf{ex}}\xspace}
\newcommand{\Iexa}{{\ensuremath{\Imc_\exa}}\xspace}
\newcommand{\Kexa}{\ensuremath{\Kmc_\exa}\xspace}
\newcommand{\Pexa}{\ensuremath{\Pmc_\exa}\xspace}
\newcommand{\Texa}{\ensuremath{\Tmc_\exa}\xspace}
\newcommand{\Wexa}{\ensuremath{\Wmc_\exa}\xspace}

\newcommand{\EL}{\ensuremath{\mathcal{E\!L}}\xspace}
\newcommand{\BEL}{\ensuremath{\mathcal{BE\!L}}\xspace}
\newcommand{\IDEL}{\ensuremath{\text{ID-}\mathcal{E\!L}}\xspace}

\newcommand{\NC}{\ensuremath{N_C}\xspace}
\newcommand{\NR}{\ensuremath{N_R}\xspace}

\newcommand{\anc}{\ensuremath{\mathsf{d\text{-}anc}}\xspace}
\newcommand{\infl}{\ensuremath{\mathsf{infl}}\xspace}
\newcommand{\val}{\ensuremath{\mathsf{val}}\xspace}
\newcommand{\cst}{\ensuremath{\mathsf{c}}\xspace}

\newcommand{\move}{\ensuremath{\mathsf{\omega}}\xspace}
\newcommand{\movez}{\ensuremath{\mathsf{\Omega}}\xspace}

\newcommand{\reals}{\ensuremath{\mathbb{R}}\xspace}

\begin{document}
%
\title{Reasoning with Contextual Knowledge and Influence Diagrams}
\author{%
Erman Acar$^1$\and
Rafael Pe\~naloza$^2$ \\
\affiliations
$^1$Vrije Universiteit Amsterdam\\
$^2$University of Milano-Bicocca\\
\emails
erman.acar@vu.nl,
rafael.penaloza@unimib.it
}

\maketitle

\begin{abstract}
Influence diagrams (IDs) are well-known formalisms extending Bayesian networks to model decision situations under 
uncertainty. Although they are convenient as a decision theoretic tool, their knowledge representation ability is limited in 
capturing other crucial notions such as logical consistency. We complement IDs with the light-weight description logic~(DL) 
\EL to overcome such limitations. We consider a setup where DL axioms hold in some contexts, yet the actual 
context is uncertain. The framework benefits from the convenience of using DL as a domain knowledge representation language 
and the modelling strength of IDs to deal with decisions over contexts in the presence of contextual uncertainty. We define related 
reasoning problems and study their computational complexity.
\end{abstract}

\section{Introduction}

A well-known limitation of classical description logics~(DLs) is their inability to deal with uncertainty~\cite{dlhandbook}. To model 
different aspects of knowledge domains where uncertainty is unavoidable, such as in the bio-medical sciences, many probabilistic extensions 
of DLs have been proposed in the literature~\cite{LuSt08,niepert2011log,GJLS-JAIR17,RBLZ15}. Among them, a prominent example are Bayesian 
DLs~\cite{CePe17,CePe-DL14,BoMP19,dAFL08}, which provide a means for expressing complex probabilistic and logical dependencies between
axioms. For example, in these logics it is easy to express that two axioms must always appear together, or that if one axiom holds, then the
likelihood of another one holding is some probability $p$.

The expressive power of Bayesian DL arises from combining a set of (classical) DL ontologies (called contexts) with a Bayesian network 
(BN)~\cite{Pear88}
representing the joint probability distribution of these ontologies. This allows to reason about the likelihood of a consequence to hold, 
given the current knowledge
and update the beliefs about the probabilities of the contexts. However, this remains a passive attitude towards knowledge,
in the sense that nothing is done with it. In
practice, an agent should be able to make choices depending on its knowledge and observations and
maximize its expected returns. BNs cannot express them.

Influence diagrams (IDs)~\cite{Shac86} generalise BNs to model potential decisions made by an agent and 
their associated costs. 
Consider for example the fictitious disease \emph{idelium}, which may remain asymptomatic, and two potential tests for 
detecting whether an individual is infected or not. Test A is cheap, but not very reliable, while Test B is much more reliable, but
expensive and intrusive. The cost of false positives and false negatives is high. The former due to the
inconveniences it causes in the life of the subject, and the latter because it can further spread the disease.
The joint probabilities of finding false positives or false negatives in the presence or absence of symptoms dependent on the
test used can be modelled via a BN. However, an agent would be more interested in deciding which
test to perform, in order to minimise the expected combined cost of test, intrusiveness, and false results. Thus, we extend
the BN to an ID which includes the decision node for the test to perform, along information about the cost of each setting
(see Figure~\ref{fig:ID}).
We propose an extension of the Bayesian DL \BEL~\cite{CePe14} 
which allows for agent decision-making combining influence diagrams with the light-weight DL \EL~\cite{BBL-EL}. We call it
\IDEL. Our main goal is to allow automated decision making in the presence of uncertainty and domain knowledge.

In \IDEL, the contexts consider the uncertainty in the network, together with the potential choices from the agent and, obviously, 
their associated costs. More importantly, the ontological knowledge can be used as evidence about the potential context, thus 
modifying the underlying probabilities, during the agent decision process.
For example, if idelium causes a green coloration of the bones, we want to add the knowledge 
$\text{Bone}\sqsubseteq\exists\text{hasColor}.\text{Green}$ which holds only in case of disease, but not when the subject
is healthy.
We study the reasoning problems associated with the 
selection, by the agent, of a strategy that minimises its expected cost given such evidence, together with other relevant
tasks.

\section{Preliminaries}

We first introduce the basic notions of influence diagrams and the DL \EL needed for 
the rest of the paper.

\subsection{Influence Diagrams}

Influence diagrams (IDs)~\cite{Shac86} are graphical models which generalise Bayesian networks (BNs)~\cite{Pear88} by 
allowing three types of
nodes: \emph{chance nodes} that reflect the uncertainty of the environment as in BNs; \emph{decision nodes}, which express
the choices made by an agent in response to the environment; and a \emph{cost node} (also called a \emph{utility node}), which
reflects the cost (or utility) of a given outcome. From a formal perspective, each of these nodes is a discrete random variable, 
and the main difference is how this variable is interpreted or used within the network.
Importantly, the agent can only influence its own decision nodes, while chance nodes can be
seen as environment attacks. 

Formally, an \emph{influence diagram} is a pair $\Dmc=(G,\Phi)$ where $G=(V\cup\{c\},E)$ is a directed acyclic graph (DAG), whose 
nodes $V$ are partitioned into two disjoint sets $B$ and $D$ of \emph{chance nodes} (or \emph{Bayesian nodes}), and 
\emph{decision nodes}, respectively,
and $c$ is a single \emph{cost node}. For simplicity, we assume w.l.o.g.\ that all nodes in $V$ are Boolean random 
variables (RVs).%
\footnote{In general, chance and decision nodes can be arbitrary finite RVs, and IDs may have more than
one cost node. Considering only Boolean RVs with a unique cost node 
greatly simplifies the notation and presentation, without affecting its generality.}
The cost node $c$ has no outgoing edges, and represents a cost function from the valuations of its parent nodes to a finite set 
$\val(c)\subseteq\reals$ of values.
For a node $v\in V\cup\{c\}$, $\pi(v)$ denotes the parents of $v$.
Given a decision node $d\in D$, $\anc(d)$ is the set of all decision ancestors of $d$, and its influence set is
\[
\infl(d):=\anc(d)\cup\pi(d). 
\]
When \Dmc uses the nodes $V\cup\{c\}$, we say that \Dmc is an ID \emph{over} $V$.
$\Phi$ is a class of conditional probability distribution tables (PDTs) $P(v\mid \pi(v))$, one for each chance
node $v\in B$ given its parents. Note that no probability distribution is associated to decision nodes, and recall that the node $c$ 
represents a function from the class of all valuations of $\pi(c)$ to $\reals$.

\begin{figure}[t]
\centering
\tikzstyle{cpt}=[rectangle,rounded corners,fill=black!20,fill opacity=0.85,text opacity=1,inner ysep=0,inner xsep=1,scale=0.95]
\begin{tikzpicture}[node distance=1.2 and 1.8,->,shorten <= 1pt,shorten >= 1pt,very thick,scale=1.3,inner sep=5]
 \node[circle,draw] (y) {\sf S};
 \node[circle,draw,above right=of y] (x) {\sf D};
 \node[rectangle,draw,below right=of y,inner sep=6] (z) {\sf TA};
 \node[circle,draw,above right=of z] (w) {\sf P};
 \node[diamond,draw,right=of w,inner sep=4] (c) {$c$};
 \draw (x) -- (y);
 \draw (x) -- (c);
 \draw (y) -- (z);
 \draw (z) -- (c);
 \draw (x) -- (w);
 \draw (z) -- (w);
 \draw (w) -- (c);
 \node[cpt,anchor=east,xshift=-1] at (x.west) {\footnotesize
 	\begin{tabular}{@{}c@{}}
	  \sf D \\ \midrule  0.3
	\end{tabular}
 };
 \node[cpt,anchor=south,xshift=-6,yshift=0] at (y.north) {\footnotesize
 	\begin{tabular}{@{}r|c@{}}
	  & \sf S \\ \midrule
	  \sf D & 0.4 \\
	  $\neg \sf D$ & 0.1
	\end{tabular}
 };
 \node[cpt,anchor=west,xshift=0,yshift=0] at (c.east) {\footnotesize
 	\begin{tabular}{@{}rrr|r@{}}
	  &&& $c$ \\ \midrule
	  \sf D & \sf P & \sf TA & 0 \\
	  \sf D & \sf P & $\neg \sf TA$ & 5 \\
	  \sf D & $\neg \sf P$ & \sf TA & 20 \\
	  \sf D & $\neg \sf P$ & $\neg \sf TA$ & 90 \\
	  $\neg \sf D$ & \sf P & \sf TA & 2 \\
	  $\neg \sf D$ & \sf P & $\neg \sf TA$ & 20 \\
	  $\neg \sf D$ & $\neg \sf P$ & \sf TA & 0 \\
	  $\neg \sf D$ & $\neg \sf P$ & $\neg \sf TA$ & 20 \\
	\end{tabular}
 };
 \node[cpt,anchor=east,text opacity=0.4,fill opacity=0.4,dashed,draw,draw opacity=0.4,xshift=-5,yshift=-2] at (z.west) {\footnotesize
 	\begin{tabular}{@{}r|c@{}}
	  & \sf TA \\ \midrule
	  \sf S & \phantom{0.7} \\
	  $\neg \sf S$ & \phantom{0.5}
	\end{tabular}
 };
 \node[cpt,anchor=east,xshift=-4,yshift=0] at (w.west) {\footnotesize
 	\begin{tabular}{@{}rr|c@{}}
	  && \sf P \\ \midrule
	  \sf D  & \sf TA & 0.7 \\
	  \sf D  & $\neg \sf TA$ & 0.9 \\
	  $\neg \sf D$ &  \sf TA & 0.4 \\
	  $\neg \sf D$ &  $\neg \sf TA$ & 0.1 \\
	\end{tabular}
 };
\end{tikzpicture}
\caption{An influence diagram; $\sf D,S,P$ are choice nodes, $\sf TA$ is a decision node, and $c$ is the cost node;
$\val(c)=\{0,2,5,20,90\}$. The probability table for the node $\sf TA$ is not specified.}
\label{fig:ID}
\end{figure}
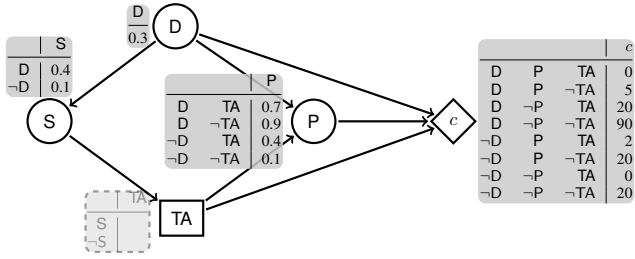

IDs are represented graphically using circles to denote chance nodes, squares for decision nodes, and a diamond for the cost
node. Figure~\ref{fig:ID} shows an ID for our fictitious idelium disease. The probability of getting an infection (\textsf{D}) is 0.3, and it has
highly specific symptoms (\textsf{S}). There are two tests: Test A (\textsf{TA}) which is cheaper, and Test B
($\neg \textsf{TA}$) which is more precise. The overall cost depends on whether the diagnosis (\textsf{P} stands for a
positive diagnosis) is correct, and which
test was used.
Seen in this way, an ID are \emph{incomplete} BN where some of the nodes are missing their conditional
probability tables, given their parents;%
\footnote{The utility function can be seen as a special kind of probability distribution over $\val(c)$, where probabilities
are always 0 or 1.}
e.g., in Figure~\ref{fig:ID}, the decision node $\sf TA$ has no associated PDT.

If the missing tables were added to the ID, then one could derive the joint probability distribution of all the
variables in $V$ using the standard chain rule from BNs
\[
P_\Dmc(V) = \prod_{v\in V} P(v\mid \pi(v)).
\]
Instead, in an ID, the decision nodes correspond to possible choices by an agent based on the information available.
The actual response of the agent is called a strategy, and each strategy has an associated value. 
Since it is the agent itself who is making the choices, these can depend on previous decisions made, of which the agent has
full knowledge. Hence, choices depend on the whole influence set of a node.

\begin{definition}[Strategy]
A (local) \emph{strategy} on a decision node $d\in D$ is a conditional PDT of $d$ given its influence set
$\infl(d)$. A (global) \emph{strategy} on the ID \Dmc is a set of local strategies, containing one for each $d\in D$. A local or global
strategy is \emph{pure} if it only assigns probabilities 0 or 1.
\end{definition}
We emphasise once again that the strategy at a decision node does not depend on its parents only, but on its whole influence set; 
that is, it depends on its decision ancestors.  
Intuitively, we can see the direction of the DAG edges as a precedence in the choices made. Hence,
every decision depends also on the choices made earlier. This can be understood as having implicit connections between the 
node $d$ and its influence set. This assumption, known as \emph{no-forgetting}, is commonly used in IDs, thus we include it in our 
formalism. However, removing it would have no effect over the results in this work, modulo a smaller size of the tables
representing local strategies.
In the ID from Figure~\ref{fig:ID}, a possible pure strategy $S$ is to assign 
$P(\textsf{TA}\mid \neg \textsf{S})=P(\neg \textsf{TA}\mid \textsf{S})=1$. 
To distinguish pure and general strategies, the former are also called \emph{actions}. 
$P_{\Dmc(S)}$ denotes the probability distribution obtained by
adopting the strategy $S$ in the ID \Dmc.

Clearly, an agent has a very large (in fact, infinite) class of strategies from which to choose. Which one is better
depends on the probability of paying
different costs given the chosen strategy. 
One usual approach for choosing a strategy is to try to minimise the expected cost.
\begin{definition}[Expected cost]
\label{def:ec}
Given a global strategy $S$ on the ID \Dmc, the \emph{expected cost} of $S$ w.r.t.\ \Dmc is
\[
E[\Dmc\mid S] := \sum_{r\in\val(c)}r\cdot P_{\Dmc(S)}(c=r).
\]
\end{definition}
The example strategy $S$ 
on the ID \Dmc of Figure~\ref{fig:ID} yields 
$P_{\Dmc(S)}(c=2)=P_{\Dmc(S)}(\neg \textsf{D},\textsf{P},\textsf{TA})=
	P_{\Dmc(S)}(\neg \textsf{D},\textsf{P},\textsf{TA},\textsf{S})=0.7\cdot 0.9\cdot 0.4=0.252$
and in general
\[
P_{\Dmc(S)}(c=r) =
	\begin{cases}
		0.504 & r=0 \\
		0.252 & r=2 \\
		0.108 & r=5 \\
		0.124 & r=20 \\
		0.012 & r=90.
	\end{cases}
\]
Hence, expected cost of this strategy is 
\begin{align*}
	E[\Dmc\mid S]= {} & 0\cdot 0.504+2\cdot 0.252+5\cdot 0.108+ {} \\ & 20\cdot 0.124+90\cdot 0.12 \\ = {} & 4.604.
\end{align*}
Strategies in IDs are often targeted to minimising the expected cost on the resulting network. However, other kinds
of problems can also be considered over these networks; e.g., finding the most likely cost, or maximising the probability of the 
minimum cost. 
If we limit ourselves to pure strategies only, then one can verify that the strategy 
$S'$ which assigns 
$P(\textsf{TA}\mid \textsf{S})=P(\textsf{TA}\mid \neg \textsf{S})=1$
maximises the probability of observing the least possible cost 0: $P_{\Dmc(S')}(c=0)=0.63$. This strategy also minimises the 
expected cost. In general, strategies reflect the response of the agent to the situations imposed by the environment.

\subsection{\EL}

\EL \cite{BBL-EL} is a light-weight description logic, which allows for polynomial reasoning in standard reasoning tasks.
As with all DLs, its main components are concepts and roles, corresponding to unary and binary predicates of first-order
logic, respectively. 

Let \NC and \NR be two disjoint sets of \emph{concept names} and \emph{role names}, respectively. \EL
\emph{concepts} are built through the grammar rule $C::= A\mid \top\mid C\sqcap C\mid \exists r.C$, where $A\in \NC$ and 
$r\in \NR$. A \emph{general concept inclusion}~(GCI) is an expression of the form $C\sqsubseteq D$, where $C,D$ are
\EL concepts, and a \emph{TBox} is a finite set of GCIs. We often call a TBox also an \emph{ontology}.
The semantics of \EL is based on \emph{interpretations}. These are tuples of the form $\Imc=(\Delta^\Imc,\cdot^\Imc)$,
where $\Delta^\Imc$ is a set called the \emph{domain} (of the interpretation) 
and $\cdot^\Imc$ is the \emph{interpretation function} which
maps every concept name $A\in\NC$ to a set $A^\Imc\subseteq \Delta^\Imc$ and every role name $r\in\NR$ to a binary
relation $r^\Imc\subseteq \Delta^\Imc\times\Delta^\Imc$. The interpretation function is extended to arbitrary \EL 
concepts by setting $\top^\Imc:=\Delta^\Imc$, $(C\sqcap D)^\Imc:=C^\Imc\cap D^\Imc$; and 
$(\exists r.C)^\Imc:=\{\delta\in\Delta^\Imc\mid\exists\gamma\in C^\Imc.(\delta,\gamma)\in r^\Imc\}$.

The interpretation \Imc \emph{satisfies} the GCI $C\sqsubseteq D$ (denoted by $\Imc\models C\sqsubseteq D$) iff 
$C^\Imc\subseteq D^\Imc$. It is a \emph{model} of the TBox \Tmc (denoted by $\Imc\models \Tmc$) iff it satisfies all
GCIs in \Tmc. Intuitively, a TBox expresses constraints on the interpretation of concepts and roles in the
knowledge domain that is being represented. Hence, we are only interested in models of the TBox. 

Since \EL cannot express negations, every TBox from this logic is consistent; i.e., it has a model.
The main reasoning problem in \EL is thus \emph{subsumption}: given a TBox \Tmc, and two \EL concepts
$C$ and $D$, $C$ is \emph{subsumed by} $D$ w.r.t.\ \Tmc ($\Tmc\models C\sqsubseteq D$) iff
every model \Imc of \Tmc satisfies the GCI $C\sqsubseteq D$. Subsumption in \EL is in PTime.

In the next section, we combine IDs with \EL, where the knowledge is divided in 
different contexts, and later study some of its reasoning problems.

\section{IDs and Contextual Ontologies }

We now introduce a new logic that combines \EL with an ID to allow 
reasoning and deriving strategies according to observed knowledge. The connection between the two formalisms is based on
adding a contextual annotation to every axiom, expressing in which circumstances it is required to hold. This notion of
a \emph{knowledge base} is formalised next.

\begin{definition}[KB]
Consider three mutually disjoint sets $V$, \NC, and \NR of \emph{contextual variables}  (or \emph{variables} for short),
\emph{concept names}, and \emph{role names}, respectively.
A (contextual) \emph{general concept inclusion}
($V$-GCI) is an expression of the form $\left<C\sqsubseteq D:\varphi\right>$ where $C,D$ are two \EL concepts and $\varphi$ is
a propositional formula over $V$. A \emph{$V$-TBox} is a finite set of $V$-GCIs. An \IDEL \emph{knowledge base} (KB) is a pair
$\Kmc=(\Dmc,\Tmc)$, where \Dmc is an ID over $V$ and \Tmc is a $V$-TBox.
\end{definition}
As with other existing context-based DLs~\cite{CePe17,BaKP12}, the idea is that a $V$-GCI is only required to hold 
when its context $\varphi$ is satisfied. This intuition is formalised via a possible world 
semantics using so-called $V$\mbox{-}interpretations. These combine
classical DL interpretations with propositional valuations to link the GCIs with their contexts.

\begin{definition}[Semantics]
A \emph{$V$-interpretation} is a triple of the form $\Imc=(\Delta^\Imc,\cdot^\Imc,\Vmc^\Imc)$, where 
$(\Delta^\Imc,\cdot^\Imc)$ is an \EL interpretation, and
$\Vmc^\Imc:V\to\{0,1\}$ is a \emph{valuation} of $V$.
The interpretation function $\cdot^\Imc$ is extended to complex concepts as usual in \EL.

The $V$-interpretation \Imc \emph{satisfies} the $V$-GCI $\left<C\sqsubseteq D:\varphi\right>$ 
($\Imc\models \left<C\sqsubseteq D:\varphi\right>$) iff $\Vmc^\Imc\not\models\varphi$ or $C^\Imc\subseteq D^\Imc$.
It is a \emph{model} of the $V$-TBox \Tmc iff it satisfies all $V$-GCIs in \Tmc.
\end{definition}
When there is no ambiguity, we omit the prefix $V$ and speak of e.g., interpretations or TBoxes.
Clearly, the probabilistic DL \BEL~\cite{CePe17}---which combines a contextual ontology with a BN---is a special case 
of \IDEL, in which there are no decision nodes, and the cost node is ignored (e.g., it may be disconnected from the rest of the 
DAG). As in that special case, it is often useful to consider the classical \EL TBoxes induced by the valuations of the 
variables in $V$\!. These correspond to the GCIs that would need to be satisfied by any model which uses this valuation.

\begin{definition}[Restricted KB]
Let $\Kmc=(\Dmc,\Tmc)$ be a KB, and \Wmc a valuation of the variables in $V$. The \emph{restriction} of \Tmc to \Wmc is the
\EL TBox
\[
\Tmc_\Wmc := \{ C\sqsubseteq D \mid \left<C\sqsubseteq D:\varphi\right>\in\Tmc, \Wmc\models\varphi \}.
\]
\end{definition}
To consider the uncertainty associated with the contexts, \BEL defines a possible world semantics where each world is 
associated with a probability that needs to be compatible with the probability distribution of the nodes. In \IDEL this
definition cannot be applied directly, because the actual probability distribution is underspecified. In fact, recall that
the full distribution depends on the strategy chosen by the agent. 
Thus, the notion of probabilistic models must be parameterised w.r.t.\ a strategy.

\begin{definition}[Probabilistic model]
A \emph{probabilistic interpretation} is a pair $\Pmc=(\Imf,P_\Imf)$, where \Imf is a finite set of $V$\mbox{-}interpretations and 
$P_\Imf$ is a probability distribution over \Imf.
This probabilistic interpretation is a \emph{model} of the TBox \Tmc if every $\Imc\in\Imf$ is a model of \Tmc.

Given an ID \Dmc and a strategy $S$ on \Dmc, the probabilistic interpretation \Pmc is \emph{consistent} with \Dmc w.r.t.\ $S$
if for every possible valuation \Wmc of the variables in $V$ it holds that 
$$P_{\Dmc(S)}(\Wmc)=\sum_{\Imc\in\Imf,\Vmc^\Imc=\Wmc}P_\Imf(\Imc).$$
\Pmc is a \emph{model} of the KB $\Kmc=(\Dmc,\Tmc)$ w.r.t.\ the strategy $S$ (denoted as $\Pmc\models_S \Kmc$) 
iff it is a model of \Tmc and consistent with \Dmc w.r.t.\ $S$.
\end{definition}
We explain these notions with a brief example.

\begin{example}
\label{exa:run}
Let $\Kexa=(\Dmc,\Texa)$ be the \IDEL KB where \Dmc is the ID in Figure~\ref{fig:ID}, and%
\footnote{Following the example in the introduction, we could add 
$\langle\text{Bone}\sqsubseteq\exists\text{hasColor}.\text{Green}: \textsf{D}\rangle$ to this TBox. We chose not to do
so to simplify the following examples.}
\begin{align*}
\Texa := \{ &
	\langle \text{Subject}\sqsubseteq \text{Infectious}: \textsf{D} \rangle,  \\ &
	\langle \text{Subject}\sqsubseteq \text{Control}: \textsf{S} \lor \textsf{P} \rangle,  \\ &
	\langle \text{Control}\sqsubseteq \text{Distance}: \textsf{S} \rangle,  \\ &
	\langle \text{Control}\sqsubseteq \text{Benefits}: \neg \textsf{P} \rangle \\ &
	\langle \text{Subject}\sqsubseteq \text{Safe}: \neg \textsf{P} \land \neg\textsf{S} \rangle &
\}.
\end{align*}
$\Wexa=\{\textsf{D},\neg\textsf{S},\textsf{TA},\neg\textsf{P}\}$ is a valuation of $V$.
The interpretation $\Iexa=(\{\delta\},\cdot^\Iexa,\Wexa)$ with
$\text{Sub}^\Iexa=\text{Inf}^\Iexa=\text{Con}^\Iexa=\{\delta\}$ and 
$\text{Dis}^\Iexa=\text{Ben}^\Iexa=\text{Saf}^\Iexa=\emptyset$
satisfies the first three GCIs, but not the last two. Indeed, $\Wexa\models \neg \textsf{P}$ but
$\text{Con}^\Iexa\not\subseteq \text{Ben}^\Iexa$, and $\Wexa\models \neg \textsf{P}\land \neg\textsf{S}$ but
$\text{Con}^\Iexa\not\subseteq \text{Saf}^\Iexa$

Let now $\Imc_i:=(\{\delta\},\cdot^{\Imc_i},\Wmc_i), 1\le i\le 8$ be the $V$\mbox{-}interpretations defined by
the interpretation functions and valuations from Table~\ref{tab:models}.
\begin{table}
\caption{Interpretation functions and valuations for Example~\ref{exa:run}.}
\label{tab:models}
\resizebox{\columnwidth}{!}{
\begin{tabular}{l*{7}{@{\hspace{0.85\tabcolsep}}c}}
$i$ & $\Wmc_i$ & $\text{Sub}^{\Imc_i}$ & $\text{Inf}^{\Imc_i}$ & $\text{Con}^{\Imc_i}$ & $\text{Dis}^{\Imc_i}$ & $\text{Ben}^{\Imc_i}$ & $\text{Saf}^{\Imc_i}$ \\ \midrule
1 & $\{\textsf{D},\textsf{S}, \textsf{P}, \neg\textsf{TA}\}$ & $\{\delta\}$ & $\{\delta\}$ & $\{\delta\}$ & $\{\delta\}$ & $\emptyset$ & $\emptyset$ \\
2 & $\{\textsf{D},\textsf{S}, \neg\textsf{P}, \neg\textsf{TA}\}$ & $\emptyset$ & $\emptyset$ & $\emptyset$ & $\emptyset$ & $\emptyset$ & $\emptyset$ \\
\rowcolor{lightgray!50}3 & $\{\textsf{D},\neg\textsf{S}, \textsf{P}, \textsf{TA}\}$ & $\{\delta\}$ & $\{\delta\}$ & $\emptyset$ & $\emptyset$ & $\emptyset$ & $\emptyset$ \\
\rowcolor{lightgray!50}4 & $\{\textsf{D},\neg\textsf{S}, \neg\textsf{P}, \textsf{TA}\}$ & $\{\delta\}$ & $\{\delta\}$ & $\emptyset$ & $\emptyset$ & $\{\delta\}$ & $\{\delta\}$ \\
5 & $\{\neg\textsf{D},\textsf{S}, \textsf{P}, \neg\textsf{TA}\}$ & $\{\delta\}$ & $\emptyset$ & $\{\delta\}$ & $\{\delta\}$ & $\emptyset$ & $\emptyset$ \\
6 & $\{\neg\textsf{D},\textsf{S}, \neg\textsf{P}, \neg\textsf{TA}\}$ & $\emptyset$ & $\emptyset$ & $\{\delta\}$ & $\{\delta\}$ & $\{\delta\}$ & $\emptyset$ \\
\rowcolor{lightgray!50}7 & $\{\neg\textsf{D},\neg\textsf{S}, \textsf{P}, \textsf{TA}\}$ & $\{\delta\}$ & $\emptyset$ & $\{\delta\}$ & $\emptyset$ & $\emptyset$ & $\emptyset$ \\
\rowcolor{lightgray!50}8 & $\{\neg\textsf{D},\neg\textsf{S}, \neg\textsf{P}, \textsf{TA}\}$ & $\emptyset$ & $\emptyset$ & $\emptyset$ & $\{\delta\}$ & $\emptyset$ & $\{\delta\}$ \\
\bottomrule
\end{tabular}
}
\end{table}
These simple interpretations are depicted in Figure~\ref{fig:exaint}.
\begin{figure}[tb]
\centering
\begin{tikzpicture}
\begin{scope}
\node[circle,draw,label=above:{Sub, Inf, Con, Dis},inner sep=2] {$\delta$};
\node[anchor=north west] at (-1.5,1.5) (i) {$\mathcal{I}_1$};
\node[anchor=south east] at (1.6,-1) (w) {$\{\textsf{D},\textsf{S},\textsf{P},\neg\textsf{TA}\}$};
\node[fit=(i)(w),rectangle,draw,rounded corners,very thick,inner ysep=0] {};
\end{scope}
\begin{scope}[shift=(0:3.45)]
\node[circle,draw,inner sep=2] {$\delta$};
\node[anchor=north west] at (-1.5,1.5) (i) {$\mathcal{I}_2$};
\node[anchor=south east] at (1.6,-1) (w) {$\{\textsf{D},\textsf{S},\neg\textsf{P},\neg\textsf{TA}\}$};
\node[fit=(i)(w),rectangle,draw,rounded corners,very thick,inner ysep=0] {};
\end{scope}
\begin{scope}[shift=(0:6.9)]
\node[circle,draw,label=above:{Sub, Inf},inner sep=2] {$\delta$};
\node[anchor=north west] at (-1.5,1.5) (i) {$\mathcal{I}_3$};
\node[anchor=south east] at (1.6,-1) (w) {$\{\textsf{D},\neg\textsf{S},\textsf{P},\textsf{TA}\}$};
\node[fit=(i)(w),rectangle,draw,rounded corners,very thick,inner ysep=0] {};
\end{scope}
\begin{scope}[shift=(-90:2.6)]
\node[circle,draw,label=above:{Sub, Inf, Ben,Saf},inner sep=2] {$\delta$};
\node[anchor=north west] at (-1.5,1.5) (i) {$\mathcal{I}_4$};
\node[anchor=south east] at (1.6,-1) (w) {$\{\textsf{D},\neg\textsf{S},\neg\textsf{P},\textsf{TA}\}$};
\node[fit=(i)(w),rectangle,draw,rounded corners,very thick,inner ysep=0] {};
\begin{scope}[shift=(0:3.45)]
\node[circle,draw,label=above:{Sub, Con, Dis},inner sep=2] {$\delta$};
\node[anchor=north west] at (-1.5,1.5) (i) {$\mathcal{I}_5$};
\node[anchor=south east] at (1.6,-1) (w) {$\{\neg\textsf{D},\textsf{S},\textsf{P},\neg\textsf{TA}\}$};
\node[fit=(i)(w),rectangle,draw,rounded corners,very thick,inner ysep=0] {};
\end{scope}
\begin{scope}[shift=(0:6.9)]
\node[circle,draw,label=above:{Con, Dis, Ben},inner sep=2] {$\delta$};
\node[anchor=north west] at (-1.5,1.5) (i) {$\mathcal{I}_6$};
\node[anchor=south east] at (1.6,-1) (w) {$\{\neg\textsf{D},\textsf{S},\neg\textsf{P},\neg\textsf{TA}\}$};
\node[fit=(i)(w),rectangle,draw,rounded corners,very thick,inner ysep=0] {};
\end{scope}
\end{scope}
\begin{scope}[shift=(-90:5.2)]
\begin{scope}[shift=(0:1.72)]
\node[circle,draw,label=above:{Sub, Con},inner sep=2] {$\delta$};
\node[anchor=north west] at (-1.5,1.5) (i) {$\mathcal{I}_7$};
\node[anchor=south east] at (1.6,-1) (w) {$\{\neg\textsf{D},\neg\textsf{S},\textsf{P},\textsf{TA}\}$};
\node[fit=(i)(w),rectangle,draw,rounded corners,very thick,inner ysep=0] {};
\begin{scope}[shift=(0:3.45)]
\node[circle,draw,label=above:{Dis, Ben},inner sep=2] {$\delta$};
\node[anchor=north west] at (-1.5,1.5) (i) {$\mathcal{I}_8$};
\node[anchor=south east] at (1.6,-1) (w) {$\{\neg\textsf{D},\neg\textsf{S},\neg\textsf{P},\textsf{TA}\}$};
\node[fit=(i)(w),rectangle,draw,rounded corners,very thick,inner ysep=0] {};
\end{scope}
\end{scope}
\end{scope}
\end{tikzpicture}
\caption{$V$-interpretations satisfying \Texa from Example~\ref{exa:run}.}
\label{fig:exaint}
\end{figure}
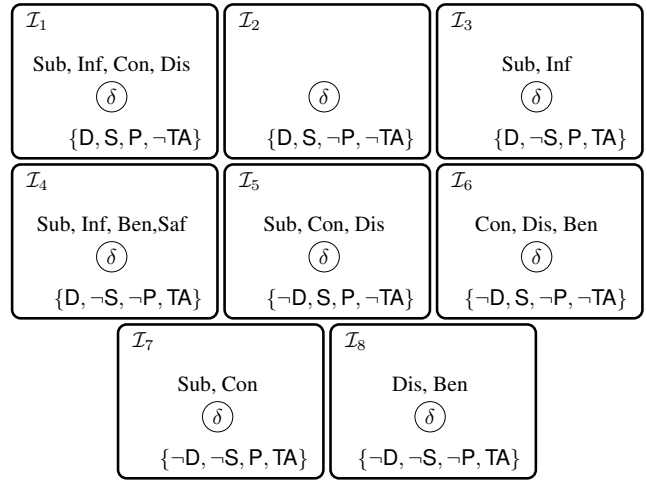
It is easy to verify that the probabilistic interpretation $\Pexa=(\Imf,P_\Imf)$ given by $\Imf=\{\Imc_1,\ldots,\Imc_8\}$ and
the distribution 
$P_\Imf$ from Table~\ref{tab:distexa} 
\begin{table}
\caption{Probability distribution for Example~\ref{exa:run}.}
\label{tab:distexa}
\resizebox{\columnwidth}{!}{
\begin{tabular}{@{}c*{8}{@{\hspace{0.65\tabcolsep}}c}@{}}
$i$ & 1 & 2 & 3 & 4 & 5 & 6 & 7 & 8 \\ \midrule
$P_\Imf(I_i)$ & 0.108 & 0.012 & 0.126 & 0.054 & 0.252 & 0.028 & 0.294 & 0.126
\end{tabular}
}
\end{table}
is a model of \Texa which is also consistent with the strategy $S$ that assigns 
$P(\textsf{TA}\mid \neg \textsf{S})=P(\neg \textsf{TA}\mid \textsf{S})=1$. Hence \Pexa is a model of \Kexa w.r.t.\ $S$.
\end{example}
In this example, we see how domain knowledge is separated from the ID. For example, we model that subjects are put under
medical control if they present symptoms or have been tested positive. Implicitly, in the context of symptoms, every subject
should keep a safe distance. Note that these are not part of the ID itself, but give us knowledge that holds in case some of
its nodes are made true.

The notion of a model is always dependent on a given strategy chosen by the agent. This is in line
with our general understanding of IDs. For instance, the strategy of an agent could be such that some contexts become 
impossible; e.g., the strategy $S$ from Example~\ref{exa:run}, requires valuations containing \textsf{S} and 
\textsf{TA} to have probability 0 (i.e., symptomatic people are always presented with test B in this strategy).
Then,
a model of the knowledge of this agent should disallow any positive probability in those contexts. As a consequence,
the basic reasoning tasks in \IDEL must also be parameterised on the chosen strategy. 
We also note that the requirement for \Imf to be finite can be relaxed by imposing some additional constraints in the 
probability distribution $P_\Imf$. To avoid unnecessary technicalities, we simply focus on the finite case.

Recall that the choice of a strategy is only a means, and the actual value of interest is the cost associated to this strategy.
We extend this idea and define the cost associated with $V$\mbox{-}interpretations and probabilistic models.
\begin{definition}[Expected cost]
Given an ID \Dmc over $V$\!, the \emph{cost} of the $V$\mbox{-}in\-ter\-pre\-ta\-tion 
$\Imc=(\Delta^\Imc,\cdot^\Imc,\Vmc^\Imc)$ is defined by
$\cst(\Imc):=c(\Vmc^\Imc|_{\pi(c)})$, where $\Vmc^\Imc|_{\pi(c)}$ denotes the restriction of the valuation $\Vmc^\Imc$ to 
the parents of $c$.

Given a strategy $S$ on \Dmc and a probabilistic interpretation $\Pmc=(\Imf,P_\Imf)$ which is consistent with \Dmc w.r.t.\ $S$, 
the \emph{expected cost} of \Pmc (w.r.t.\ $S$) is $$E[\Pmc\mid S]:=\sum_{\Imc\in\Imf}P_\Imf(\Imc)\cdot\cst(\Imc).$$
\end{definition}%
Since the probability distribution in a probabilistic model must be consistent with the distribution induced by the strategy $S$,
the expected cost of any model of a KB $\Kmc=(\Dmc,\Tmc)$ w.r.t.\ $S$ corresponds exactly to the expected cost of \Dmc w.r.t.\
$S$. That is, once that the strategy has been chosen, the expected cost does not depend on the specific model of \Kmc,
but only on the probabilities associated to this strategy. Thus, we can define the
\emph{expected cost} of the KB \Kmc w.r.t.\ $S$ as $E[\Kmc\mid S]:=E[\Pmc\mid S]$, where \Pmc is any model of \Kmc.

Before moving to the next section, we present the following remark.
Rather than defining a cost function directly on the nodes of the network, it sometimes makes sense to consider this function
to be implicitly defined by the properties of the contexts that the node $c$ can observe. In the extreme case, all nodes in
$V$ are parents of $c$ and defining the cost function in terms of the contexts obtained by each valuation avoids having to
represent the exponentially large mapping. 
A natural choice for such a cost function is the size of the context. Intuitively, this function would allow us to express that a smaller
context is preferred over a larger one. Using this function makes sense, for instance, when the context needs to be transferred
or manipulated over an unreliable channel. A smaller ontology is preferred to reduce the risk of errors. However, many other
functions can be considered depending on the application. As an additional example, if the contexts refer to different levels
of granularity of access, then considering the size of the vocabulary as cost is more relevant, as a larger vocabulary 
corresponds to a wider access to the knowledge.
We emphasise, however, that \IDEL does not require the use of any of these cost functions, or even that the node $c$ is
influenced by all nodes in $V$\!. These are just given as concrete examples with an application-oriented motivation.

\section{Reasoning in \IDEL}

Before delving into the reasoning tasks for \IDEL, note that as in the special cases \EL and \BEL, every
\IDEL ontology is consistent: for every \IDEL KB and every strategy $S$, there is a model of \Kmc w.r.t.\ $S$,
which can be built as follows.
Let $\Kmc=(\Dmc,\Tmc)$ be a KB and $S$ a strategy, and let $V$ be the set of variables in \Dmc. For every valuation 
\Wmc of $V$, consider the interpretation $\Imc_\Wmc=(\{\delta\},\cdot^{\Imc_\Wmc},\Wmc)$ where 
$A^{\Imc_\Wmc}=\emptyset$ and $r^{\Imc_\Wmc}=\emptyset$ for all $A\in\NC, r\in\NR$. The probabilistic
interpretation $\Pmc=(\Imf,P_\Imf)$ with $\Imf=\{\Imc_\Wmc\mid \Wmc$ is a valuation of $V\}$ and
$P_\Imf(\Imc_\Wmc)=P_{\Dmc(S)}(\Wmc)$ is a model of \Kmc w.r.t.\ $S$.
Hence, we are more interested in reasoning problems related to subsumptions (as in \EL), their probabilities
(as in \BEL), but most importantly, their costs.

The first reasoning task that we consider in this setting corresponds to computing bounds on the expected costs associated
with the models of a given KB \Kmc. To this end, we would like to find an \emph{optimal strategy}, which minimises the 
expected cost w.r.t.\ \Dmc, and a \emph{pessimal strategy}, which maximises this cost. 
From the previous discussion, it follows that these bounds correspond exactly to the bounds on the expected cost of the ID \Dmc 
from \Kmc. 
In order to study the computational complexity of finding these bounds, we consider their associated decision problem versions.
\begin{problem}[Optimal/Pessimal strategy]
Consider an ID \Dmc and a value $b\in\reals$. The \emph{optimal strategy problem} (D-Opt) consists in deciding whether there 
exists a strategy $S$ such that $E[\Dmc\mid S] < b$. Dually, the \emph{pessimal strategy problem} (D-Pes) is to decide whether 
there exists a strategy $S$ such that $E[\Dmc\mid S] > b$.
\end{problem}
As both problems are \textsc{PSpace}-complete~\cite{LiMP01},
their extension to the setting of \IDEL KBs,
must also be \textsc{PSpace}-complete.
\begin{theorem}
\label{thm:psp}
Given an \IDEL KB $\Kmc=(\Dmc,\Tmc)$ and $b\in\reals$, deciding if there exists a strategy $S$ such that 
$E[\Kmc\mid S]<b$ (or, dually, such that $E[\Kmc\mid S]>b$) is \textsc{PSpace}-complete.
\end{theorem}
However, in general we cannot expect a polynomial-space algorithm
to enumerate an optimal strategy. Indeed, even if we limit the search to pure strategies (which means that the probability
tables for decision nodes is Boolean), every pure local strategy corresponds to
a Boolean function over the parent variables. It is well known that for every $n\ge 2$ there are Boolean functions (and hence,
local strategies) that cannot be expressed with circuits of size smaller or equal to $2^n/2n$~\cite{Shan49}. It is not hard
to construct, for a Boolean function $f$, an ID whose optimal pure strategy is in fact $f$, which translates the hardness
result to our setting.

One can also consider the problem of entailment of a contextual subsumption, or computing the probability of a subsumption
relation to hold. For the latter, as already explained, one must first instantiate the chosen strategy, as otherwise the
probability is not well-defined.

\begin{definition}[Probabilistic subsumption]
Let $\Kmc=(\Dmc,\Tmc)$ be a KB, $\alpha$ a context, and $C,D$ two \IDEL concepts.
Given the probabilistic interpretation $\Pmc=(\Imf,P_\Imf)$, the \emph{probability} of $\langle C \sqsubseteq D: \alpha \rangle$ w.r.t.\ \Pmc 
and w.r.t.\ the strategy $S$ over \Dmc are defined, respectively, as 
\begin{align*}
P(\langle C \sqsubseteq_\Pmc D :\alpha \rangle):= {} & 
	\sum_{\Imc\in \Imf, \Imf\models \langle C \sqsubseteq D: \alpha \rangle} P_\Imf(\Imc),  & \text{and}\\
P(\langle C \sqsubseteq_{\Kmc,S} D: \alpha \rangle):= {} & \inf_{\Pmc \models_S \Kmc} P(\langle C \sqsubseteq_\mathcal{P} D :\alpha \rangle).
\end{align*}
\end{definition}
In particular, we denote as $P(C\sqsubseteq_\Pmc D)$ the case where $\alpha={\sf true}$ is the universal context satisfied by all 
propositional valuations. Iin this case, satisfaction of an axiom by an interpretation corresponds exactly to the
classical definition in \EL, as the condition of violating the context can never hold.

Recall that an ID together with a strategy forms a BN, and hence after choosing the strategy, the probability of each instantiation
of all the variables in $V$ is fully specified. Still, one can choose different models for the KB w.r.t.\ this strategy. Indeed, note
that the universal \EL model which contains only one element belonging to all concepts and connected to
itself via all roles, or the empty model defined at the beginning of this section, can always be used to build a probabilistic model 
\Pmc such that 
$P(\langle C \sqsubseteq_\Pmc D :\alpha \rangle)=1$ for all concepts $C,D$ and contexts $\alpha$. 
Choosing the infimum in the definition of the
probability of a subsumption is the natural cautious bound that is guaranteed to hold in \emph{all} models.

In a decision situation, an agent might observe a fact, and try to act upon it with the best available strategy. In IDs, the 
observations made are modelled
through the introduction of \emph{evidence}, which formally is just the instantiation of one (or more) of the chance nodes. 
In our setting, we 
are more interested in observing facts that arise from the ontological perspective. That is, our knowledge about the 
context is not directly accessible through the variables of the ID, but rather through the consequences that are known
to hold. 

Hence, rather than observing the behaviour
of the ID, we observe a fact that provides information about the possible contexts that can still hold. This information
obviously also influences the probability distribution over the underlying ID, even if the truth value of all nodes in the 
graph may still remain uncertain.
In practice, when we observe a consequence, we can immediately exclude some cases (i.e., contexts) which contradict our 
observation. The probabilities of the remaining cases need to be updated accordingly, after the impossible cases are removed.
%

\begin{definition}[Conditional expected cost]
Let $\Kmc=(\Dmc,\Tmc)$ be an \IDEL KB, $S$ a strategy on \Dmc, $\Pmc=(\Imf,P_\Imf)$ a probabilistic model of \Dmc w.r.t.\ $S$, and $C,D$ 
two concepts such that $P(C\sqsubseteq_\Pmc D)>0$. 
The \emph{conditional probability} of the interpretation $\Imc\in\Imf$ given the subsumption $C\sqsubseteq D$ is
\[
P_\Imf(\Imc\mid C\sqsubseteq D) :=
	\begin{cases}
		0 & \text{if $\Imc\not\models C\sqsubseteq D$} \\
		\frac{P_\Imf(\Imc)}{P(C\sqsubseteq_\Pmc D)} & \text{otherwise.}
	\end{cases}
\]
The \emph{conditional expected cost} of \Pmc given $C\sqsubseteq D$ w.r.t.\ $S$ is 
\begin{align*}
E[\Pmc\mid S,C\sqsubseteq D] := {} & \sum_{\Imc\in\Imf}P_\Imf(\Imc\mid C\sqsubseteq D)\cdot c(\Imc).
\end{align*}
\end{definition}
Like when dealing with probabilities alone, if one is trying to understand the expected cost given an observation 
it becomes important to consider all the possible models of the KB. Accordingly, we can consider an optimistic or a pessimistic 
approach depending on whether we try to maximise or minimise this expected cost.

\begin{definition}
Let \Kmc be an \IDEL KB, $S$ a strategy, and $C,D$ two concepts. The \emph{optimistic expected cost} $\underline{E}$ and the 
\emph{pessimistic expected cost} $\overline{E}$ of \Kmc w.r.t.\ $S$ given $C\sqsubseteq D$ are defined, respectively, by
\begin{align*}
\underline{E}[\Kmc\mid S,C\sqsubseteq D] := {} & \inf_{\Pmc\models_S\Kmc} E[\Pmc\mid S,C\sqsubseteq D], \\
\overline{E}[\Kmc\mid S,C\sqsubseteq D] := {} & \sup_{\Pmc\models_S\Kmc} E[\Pmc\mid S,C\sqsubseteq D].
\end{align*}
\end{definition}
Note that, as mentioned already, for every context it is always possible to construct an \EL model of the context that satisfies also 
the GCI $C\sqsubseteq D$. In the models $\Pmc=(\Imf,P_\Imf)$ constructed as explained earlier in this paper, it always holds that 
$P_\Imf(\Imc\mid C\sqsubseteq D)=P_\Imf(\Imc)$ for all $\Imc\in\Imf$. In particular, this also means that 
$E[\Pmc\mid S,C\sqsubseteq D]=E[\Pmc\mid S]$ for all strategies $S$. This yields the following result.

\begin{proposition}
\label{prop:bounds}
For every \IDEL KB $\Kmc$, strategy $S$, and concepts $C,D$, it holds that 
$$\underline{E}[\Kmc\mid S,C\sqsubseteq D] \le E[\Kmc\mid S] \le \overline{E}[\Kmc\mid S,C\sqsubseteq D].$$
\end{proposition}
This proposition means that the expected cost of a KB w.r.t.\ a given strategy---which, as seen earlier, 
corresponds to the expected cost of its underlying ID---provides bounds for the expected costs given an observed 
consequence. This can immediately help prune the search space for the optimistic and pessimistic bounds. In addition,
it hints at the idea that these bounds can be found by manipulating the distribution of the ID.
As the following example shows, the inequalities from Proposition~\ref{prop:bounds} may be strict.
\begin{example}
Consider again the KB \Kexa from Example~\ref{exa:run}.
We have already seen that under the pure strategy $S$ defined by 
$P(\textsf{TA}\mid \neg \textsf{S})=P(\neg \textsf{TA}\mid \textsf{S})=1$, it holds that 
$$E[\Kexa\mid S]=E[\Dmc\mid S]=4.604.$$
Consider the
evidence $\text{Subject}\sqsubseteq \text{Benefits}$. 
Figure~\ref{fig:exaexp} depicts $V$-interpretations, which form a probabilistic model \Pmc of \Kexa with the 
same probability distribution as in Example~\ref{exa:run}. 
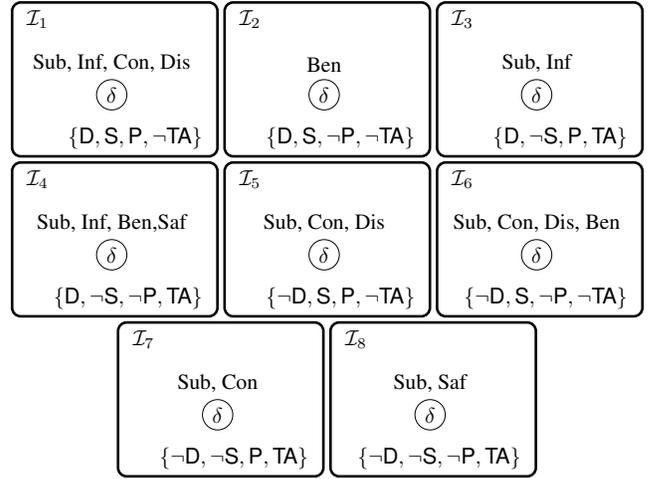
\begin{figure}[tb]
\centering
\begin{tikzpicture}
\begin{scope}
\node[circle,draw,label=above:{Sub, Inf, Con, Dis},inner sep=2] {$\delta$};
\node[anchor=north west] at (-1.5,1.5) (i) {$\mathcal{I}_1$};
\node[anchor=south east] at (1.6,-1) (w) {$\{\textsf{D},\textsf{S},\textsf{P},\neg\textsf{TA}\}$};
\node[fit=(i)(w),rectangle,draw,rounded corners,very thick,inner ysep=0] {};
\end{scope}
\begin{scope}[shift=(0:3.45)]
\node[circle,draw,label=above:{Ben},inner sep=2] {$\delta$};
\node[anchor=north west] at (-1.5,1.5) (i) {$\mathcal{I}_2$};
\node[anchor=south east] at (1.6,-1) (w) {$\{\textsf{D},\textsf{S},\neg\textsf{P},\neg\textsf{TA}\}$};
\node[fit=(i)(w),rectangle,draw,rounded corners,very thick,inner ysep=0] {};
\end{scope}
\begin{scope}[shift=(0:6.9)]
\node[circle,draw,label=above:{Sub, Inf},inner sep=2] {$\delta$};
\node[anchor=north west] at (-1.5,1.5) (i) {$\mathcal{I}_3$};
\node[anchor=south east] at (1.6,-1) (w) {$\{\textsf{D},\neg\textsf{S},\textsf{P},\textsf{TA}\}$};
\node[fit=(i)(w),rectangle,draw,rounded corners,very thick,inner ysep=0] {};
\end{scope}
\begin{scope}[shift=(-90:2.6)]
\node[circle,draw,label=above:{Sub, Inf, Ben,Saf},inner sep=2] {$\delta$};
\node[anchor=north west] at (-1.5,1.5) (i) {$\mathcal{I}_4$};
\node[anchor=south east] at (1.6,-1) (w) {$\{\textsf{D},\neg\textsf{S},\neg\textsf{P},\textsf{TA}\}$};
\node[fit=(i)(w),rectangle,draw,rounded corners,very thick,inner ysep=0] {};
\begin{scope}[shift=(0:3.45)]
\node[circle,draw,label=above:{Sub, Con, Dis},inner sep=2] {$\delta$};
\node[anchor=north west] at (-1.5,1.5) (i) {$\mathcal{I}_5$};
\node[anchor=south east] at (1.6,-1) (w) {$\{\neg\textsf{D},\textsf{S},\textsf{P},\neg\textsf{TA}\}$};
\node[fit=(i)(w),rectangle,draw,rounded corners,very thick,inner ysep=0] {};
\end{scope}
\begin{scope}[shift=(0:6.9)]
\node[circle,draw,label=above:{Sub, Con, Dis, Ben},inner sep=2] {$\delta$};
\node[anchor=north west] at (-1.5,1.5) (i) {$\mathcal{I}_6$};
\node[anchor=south east] at (1.6,-1) (w) {$\{\neg\textsf{D},\textsf{S},\neg\textsf{P},\neg\textsf{TA}\}$};
\node[fit=(i)(w),rectangle,draw,rounded corners,very thick,inner ysep=0] {};
\end{scope}
\end{scope}
\begin{scope}[shift=(-90:5.2)]
\begin{scope}[shift=(0:1.72)]
\node[circle,draw,label=above:{Sub, Con},inner sep=2] {$\delta$};
\node[anchor=north west] at (-1.5,1.5) (i) {$\mathcal{I}_7$};
\node[anchor=south east] at (1.6,-1) (w) {$\{\neg\textsf{D},\neg\textsf{S},\textsf{P},\textsf{TA}\}$};
\node[fit=(i)(w),rectangle,draw,rounded corners,very thick,inner ysep=0] {};
\begin{scope}[shift=(0:3.45)]
\node[circle,draw,label=above:{Sub, Saf},inner sep=2] {$\delta$};
\node[anchor=north west] at (-1.5,1.5) (i) {$\mathcal{I}_8$};
\node[anchor=south east] at (1.6,-1) (w) {$\{\neg\textsf{D},\neg\textsf{S},\neg\textsf{P},\textsf{TA}\}$};
\node[fit=(i)(w),rectangle,draw,rounded corners,very thick,inner ysep=0] {};
\end{scope}
\end{scope}
\end{scope}
\end{tikzpicture}
\caption{Interpretations forming a model \Pmc of \Kexa such that 
$\overline{E}[\Kexa\mid S,A\sqsubseteq C]=E[\Pmc\mid S,A\sqsubseteq C]$ and
$\underline{E}[\Kexa\mid S,A\sqsubseteq D]=E[\Pmc\mid S,A\sqsubseteq D]$.}
\label{fig:exaexp}
\end{figure}
We check that $E[\Pmc\mid S,\text{Sub}\sqsubseteq \text{Ben}]=60.149$.
Note that $\Imc_2,\Imc_4$, and $\Imc_6$ entail $\text{Sub}\sqsubseteq \text{Ben}$, but all other interpretations
do not entail it. Hence, $P(\text{Sub}\sqsubseteq_\Pmc \text{Ben})=P_\Imf(\Imc_2)+P_\Imf(\Imc_4)+P(\Imc_6)=0.094$, and the
conditional probabilities are
\[
P_\Imf(\Imc_i\mid \text{Sub}\sqsubseteq \text{Ben}) =
	\begin{cases}
	  \frac{0.012}{0.094} & i=2 \\
	  \frac{0.054}{0.094} & i=4 \\
	  \frac{0.028}{0.094} & i=6 \\
	  0 & \text{otherwise.} 
	\end{cases}
\]
Thus $E[\Pmc\mid S,\text{Sub}\sqsubseteq \text{Ben}]=90\frac{12}{94}+90\frac{54}{94}+20\frac{28}{94}=69.149$. 
It is easy to verify that $\overline{E}[\Kexa\mid S,A\sqsubseteq C]=69.149$. The worst-case scenario
is that we get a cost of 90 (the highest possible in our model) with probability 1, thus giving a expected cost of 90; however,
our evidence is such that every model satisfying the valuation $\neg\textsf{D},\neg\textsf{P}$ must also satisfy 
$\text{Sub}\sqsubseteq\text{Ben}$. Since the cost associated to this valuation is 20, the overall expected cost must decrease,
as in our model.

Similarly, $E[\Pmc\mid S,\text{Sub}\sqsubseteq \text{Saf}]=1.5$. The only 
$V$-interpretations that satisfy 
$\text{Sub}\sqsubseteq \text{Saf}$ are $\Imc_4$ and $\Imc_8$, which are associated to cost 5 and 0, respectively,
which yields $E[\Pmc\mid S,\text{Sub}\sqsubseteq \text{Saf}]=5\frac{54}{180}=1.5$.
It follows that 
$\underline{E}[\Kexa\mid S,\text{Sub}\sqsubseteq \text{Saf}]\le 1.5$.

Hence, in general the pessimistic and optimistic 
expected costs given an evidence do not coincide with the expected cost of the KB. This example also shows that different 
models may reduce or increase the expected cost, in manners that may not be obvious at first sight.
\end{example}
This example suggests a method for computing the optimistic and pessimistic expected costs. For the former,
we want to maximize the probability of observing the smallest possible costs, while minimizing the probability of getting high
costs. The dual approach helps us find the pessimistic expected cost.

\begin{theorem}
\label{thm:exp}
Optimistic and pessimistic expected costs given $C\sqsubseteq D$ can be computed in polynomial space on the number of 
nodes of the underlying ID.
\end{theorem}
%
\begin{proof}
There are exponentially many valuations of the variables in $V$, but each of them is linearly represented in the size of $V$. 
For each valuation \Wmc, we construct the TBox $\Tmc_\Wmc$.
Let $n$ be the smallest value in $\val(c)$. We construct a probabilistic model $\Pmc=(\Imf,P_\Imf)$ as follows.
For each valuation \Wmc, \Imf contains a $V$-interpretation $\Imc_\Wmc=(\Delta^{\Imc_\Wmc},\cdot^{\Imc_\Wmc},\Wmc)$ 
such that (i) $(\Delta^{\Imc_\Wmc},\cdot^{\Imc_\Wmc})\models \Tmc_\Wmc$, (ii)~if $\cst(\Imc)=n$ then 
$(\Delta^{\Imc_\Wmc},\cdot^{\Imc_\Wmc})\models C\sqsubseteq D$, and (iii)~if $\cst(\Imc)\not=n$ and 
$\Tmc_\Wmc\not\models C\sqsubseteq D$, then $(\Delta^{\Imc_\Wmc},\cdot^{\Imc_\Wmc})\not\models C\sqsubseteq D$.
Moreover, $P_\Imf(\Imc_\Wmc)=P_{\Dmc(S)}(\Wmc)$. It is easy to verify that this is a model, constructed in exponential time,
which minimises the expected cost. To compute this cost in polynomial space, we store only one interpretation at a time, and 
accumulate the relative cost of each interpretation iteratively.
For the pessimistic expected cost, the proof is analogous, but using the largest value of $\val(c)$ instead.
\end{proof}
We are not interested in the expected costs \emph{per se}, but rather as a means to identify the \emph{best} strategy for 
the agent to follow under the evidence. We thus have the choices to minimise or maximise the optimistic or pessimistic 
expected costs, yielding four different notions. To reduce the overhead of the definition, we focus only on minimising these 
costs; maximisation can be treated analogously, with just the obvious modifications in the definitions
and techniques.
\begin{definition}[Dominant strategies]
\label{def:os}
Let \Kmc be an \IDEL KB and $C,D$ two concepts. The strategy $S$ is called \emph{dominant optimistic} if for every strategy $S'$ it 
holds that
$$\underline{E}[\Kmc\mid S,C\sqsubseteq D]\le\underline{E}[\Kmc\mid S',C\sqsubseteq D].$$ 
It is \emph{dominant pessimistic} if for all strategies $S'$, 
$$\overline{E}[\Kmc\mid S,C\sqsubseteq D]\le\overline{E}[\Kmc\mid S',C\sqsubseteq D].$$
\end{definition}
To avoid confusions, we emphasise that a dominant pessimistic strategy \emph{minimizes} the pessimistic
expected cost. In terms of decision making, such a strategy ensures that in the worst case, the overall cost remains
manageable.

A na\"ive approach for finding \emph{pure} dominant strategies is to enumerate all possible options, 
bulding the Boolean functions for each local strategy, and preserving those that yield the 
lowest expected costs. 
In the worst case, there are doubly-exponentially many such strategies on the size of $V$, which makes this na\"ive approach 
infeasible, despite its effectiveness. 
On the other hand, it is easy to see that the optimal strategy for the whole network is a special case 
of Definition~\ref{def:os}, where the subsumption $C\sqsubseteq D$ of interest corresponds to any \EL tautology 
(e.g., $A\sqsubseteq A$). 

Consider the decision problems (D-Dom-Opt and D\mbox{-}Dom-Pes, respectively) associated with Definition~\ref{def:os}: given a KB \Kmc, two 
concepts $C,D$ and $b\in\reals$, decide whether there are strategies $S,S'$ such that $\underline{E}[\Kmc\mid S,C\sqsubseteq D]<b$, and 
$\overline{E}[\Kmc\mid S',C\sqsubseteq D]<b$, respectively.
Using an approach similar to Theorem~\ref{thm:exp}, we can build a polynomial-space algorithm for deciding D-Dom-Opt 
under pure strategies by
enumerating all valuations of the chance nodes, guessing for each of them a valuation of the decision variables and computing
the minimal cost that arises from each of them. The only issue is that this needs to be done in a specific order to guarantee that
for equal parent nodes, the same guess is made always in a decision variable.

\begin{theorem}
The problems D-Dom-Opt and D-Dom-Pes are {\sc PSpace}-complete for pure strategies.
\end{theorem}
\begin{proof}[Proof sketch]
{\sc PSpace}-hardness follows from Theorem~\ref{thm:psp} since D-Opt is a special case of D-Dom-Opt. For the upper bound,
we use the result from Theorem~\ref{thm:exp}: to verify D-Opt, for every valuation of the chance variables ($B$), we can
guess (in polynomial time) a valuation of the decision variables ($D$) and compute in polynomial space its expected cost. This gives
a non-deterministic polynomial space algorithm, which by Savitch's theorem \cite{Savi70} yields a {\sc PSpace} upper bound.
\end{proof}
Obviously, the {\sc PSpace} complexity lower bound holds also for arbitrary strategies, as it is a more general problem. 
The upper bound can be extended to non-pure strategies,
as long as they are representable in exponential space; otherwise, we would not be able to guess them in exponential time.
The biggest problem when dealing with arbitrary strategies is that there are uncountably many of them, and a different choice 
in one decision node may greatly affect the probability in a node that it influences. 

\subsection*{Computing Optimal Arbitrary Strategies}

We have previously restricted ourselves to pure strategies. We now remove this restriction, and investigate the case where the 
strategies can be arbitrary; namely, mixed strategies which generalizes the notion of pure strategy. To carry out our analysis, we 
incorporate techniques from game theory; these will be \emph{sequential forms} in \emph{extended-form} games 
\cite{Fudenberg91,NisaRougTardVazi07}. In what follows, we assume basic knowledge of game theory (see \cite{Fudenberg91} for further details). 

As mentioned already, an influence diagram can be understood as an agent making decisions against nature. In game-theoretic 
terms, this can be thought of as a two-player game where the nodes of the influence diagram are partitioned in two sets which 
belonging to the different players. In particular, one player (which we call the \emph{first player} or \emph{Player 1}), is the agent 
endowed with the decision nodes and the cost function (implemented by our cost node) and the other player (the \emph{second 
player}) is nature endowed with chance nodes.  Player 1 has a preference between different outcomes determined by the 
(expected) cost while Player 2 is indifferent between any outcome.%
\footnote{This can be implemented by any cost function whose codomain is a singleton.} 
This interpretation will help us make use of any formal tool which is used to find (Nash) equilibria (i.e., a state where both players 
minimize their cost function concurrently. The reason is that, in such a setting, the equilibrium notion boils down to the case where it 
is dependent only on a single player, namely the case in which Player 1 minimizes their cost function.  

To incorporate the techniques from extended-form games for a given influence diagram $\mathcal{D}$, we construct a game tree 
$\mathcal{G}_\mathcal{D}=(V_g, E_g)$ from $\mathcal{D}$. As this process is quite intuitive, to avoid cumbersome technicalities, we give a rather brief and informal description: 
\begin{enumerate}
\item[$(i)$]  for every chance and decision node in the ID, add a node and label it by the player that controls it i.e., 1 if it is a decision node and 2 if it is a chance node;
\item[$(ii)$] add a directed edge for every value of that node where the source is the player which controls the value of that node  and the target is the node (player) which controls the child node in the ID; and
\item[$(iii)$] add a leaf node for every value of the cost node respecting the path in the ID.
\end{enumerate} 
For simplicity, we assume w.l.o.g.\ that players are alternating. Dropping this assumption does not affect our analysis, since it can 
easily be converted to such a form: if several nodes of the same type are consecutive, they can be replaced by a single
non-binary random variable.

\begin{figure}[tb]
    \centering
  \includegraphics[scale=0.1255]{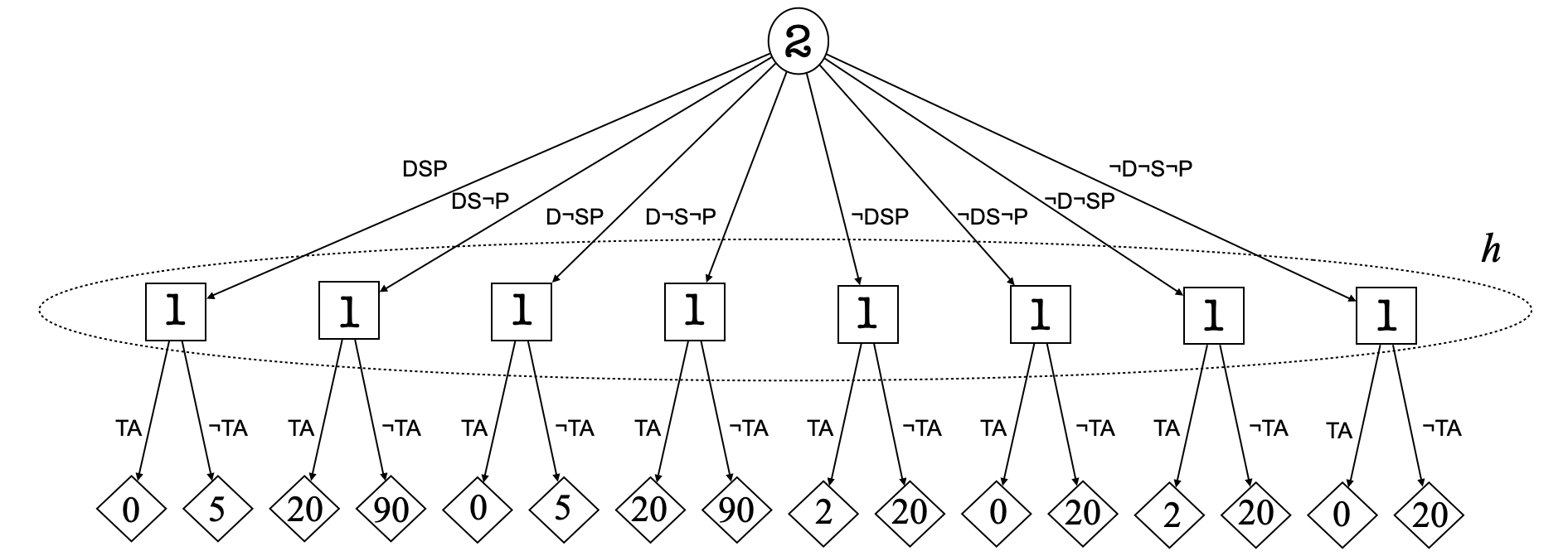}
    \caption{The game tree obtained from Figure~\ref{fig:ID}.}
    \label{fig:decisiontree}
\end{figure}

Consider for example the game tree obtained by transforming the influence diagram in Figure~\ref{fig:ID}, which
is depicted in Figure~\ref{fig:decisiontree}.
Observe that every non-terminal node is labeled by the player which controls it in the influence diagram. Moreover, shapes of the 
node for players correspond to the shape of the nodes (i.e., 1 and 2) that they control in the ID. In parallel, leaf nodes 
$\ell \in \mathcal{L}$ which represent outcomes are diamond-shaped. 
Since the root node in the ID (in Figure~\ref{fig:ID}), player 2 (nature) is the root node in the game tree. 

We introduce some necessary notions for computing the cost minimization problem with arbitrary strategies. A \emph{move} 
$\move$ is an edge in the game tree, and corresponds to a value of a particular decision node in an ID i.e., $\move \in val(d)$. 
A \emph{sequence} $\sigma$ is a path (a sequence of moves) in the game tree. For example, the empty sequence $\emptyset$ 
and  $\textsf{DS}\neg\textsf{PTA} $  are sequences in Figure~\ref{fig:decisiontree}. A sequence of player $i$ is the sequence $\sigma_i$ of its moves 
along the game tree e.g., $\textsf{DS}\neg\textsf{P} $  for $\sigma_2$ and $\textsf{TA}$ for $\sigma_1$. We denote all possible sequences of Player 1 and 
Player 2, by $\mathcal{S}_1$ and $\mathcal{S}_2$, respectively. Given a leaf node  $\ell \in \mathcal{L}$, we denote the  
sequence which reaches $\ell$ by $\sigma(\ell)$, and $\sigma_i(\ell)$ denotes the sequential moves of Player $i$. We refer to 
the content of such a leaf, as $c(\ell) \in val(c)$, after the cost function $c$.%
\footnote{Traditionally, one would write $c_1(\ell)$ to denote the cost of Player 1 in $\ell$. We drop this since we are interested 
in a single player.}   

An \emph{information set} is a set of nodes in which a player has the same moves e.g., oval $h$ in Figure~\ref{fig:decisiontree}. 
In our case, we collapse information sets to singletons since Player 1 (the agent in ID) has access to the conditional probability 
distribution of each chance node. In game-theoretic terms, this correspond to \emph{perfect information game} where Player 1 
can observe what Player 2 has chosen. We denote available moves for a player $i$ in an information set $h$ as 
$\movez_h$ which corresponds to values of a particular node $d$ in the ID, and the set of its information sets as $H_i$.
If both players can remember all of their moves along the path of the game tree, the game has \emph{perfect recall}. Intuitively, 
it means that no player can get additional information about their position in the game three by remembering earlier actions. 
Observe that this is indeed inline with our initial assumption of \emph{no forgetting}.

We translate the conditional distributions in the ID to the game tree  in a way that it simulates the overall behaviour faithfully. 
We use the well-known notion of \emph{behaviour strategy} in extended-form games. A behaviour strategy is a probability 
distribution $\beta$ on the next available moves in a state in the game tree.  For instance, in Figure~\ref{fig:decisiontree}, the 
probability that the move  $\textsf{DS}\neg\textsf{P}$  is taken can be chosen w.r.t.\ the ID in Figure~\ref{fig:ID} and the strategy $S$, which would be, then, 0.012. Hence $\beta_2(\textsf{DS}\neg\textsf{P}) = 0.012$ where $\beta_2$ is the behavior strategy for Player 2. 
Obviously, a behaviour strategy $\beta_i$ for player $i$ satisfies the following: $\sum_{\move \in \movez_h} \beta_i(\move)=1$ 
and $\beta_i(\move)\geq 0$ for all $h\in H_i $, $\move \in \movez_h$. 
Then by extending behaviour strategies to sequences, we obtain \emph{realization probability} of a sequence $\sigma$ of 
Player $i$ under $\beta_i$:  $\beta_i(\sigma) := \Pi_{\move \in \sigma } \beta_i(\move)$
which is in line with the standard chain rule.  Note that $\beta_i(\emptyset)=1$ for any $\beta_i$. In terms of the game tree, 
the expected cost (in Definition~\ref{def:ec}) can be rewritten as 
\begin{equation} 
  \sum_{\ell \in \mathcal{L}} c(\ell) \cdot \beta_1(\ell) \cdot \beta_2(\ell).\label{eq:gamecost}
\end{equation}
Moreover, given a $h \in H$  and a sequence $\sigma \move$ being an extension of sequence $\sigma \in h$ with a move 
$\move$, then we define
\begin{equation}
 \beta_i(\sigma):= \Sigma_{\move \in h} \beta_i (\sigma \move).\label{eq:realseq}
\end{equation} 

We define $\beta_i(\sigma') = 0$ for any non-realizable sequence $\sigma'$. Then any $\mu_1$ being  a vector value of such a 
$\beta_1$ (hence for Player 1) is called a \emph{realization plan}.%
\footnote{This extension of $\beta$  is a mixed-strategy (over  $\mathcal{S}_1$). By Kuhn's theorem 
\cite{maschler2013game}, if a player has perfect recall, then a mixed strategy is equivalent to a behaviour strategy.} 
This is analogous for Player 2, but since it is indifferent for any outcome, and observing that its realization plan is  fixed,%
\footnote{Indeed, this is the case since given an ID, chance nodes have certain distributions which can be considered as a fixed 
global strategy realizing $\mu_2$.}
considering only the expected cost minimization of Player 1 fulfills our goals. We can represent the cost of player 1 in terms of a 
\emph{cost matrix} $C$, as follows. For every leaf node $\ell \in \mathcal{L}$ in $\mathcal{G}_\mathcal{D}$, the entries
$
  c_{\sigma_1(\ell) \sigma_2(\ell)} := c(\ell)
$
construct a $|\mathcal{S}_1| \times |\mathcal{S}_2|$ matrix. The expected cost of Player 1 is
$ \mu_1^{\top} (C \mu_2) $ where 
 $\mu_1$ is the realization plan of Player 1 (a global strategy),  $C$ is the cost matrix, and $\mu_2$ is the realization plan of nature 
 (chance nodes).

We now have all we need to formulate the expected cost minimization problem of Player 1 in terms of linear constraints. Given a fixed realization plan $\mu_2$,
\begin{equation}
    \min \mu_1(C \mu_2) \text{ subject to }R \cdot \mu_1=r,\text{ } \mu_1\geq 0 \label{eq:LP}
\end{equation}
where $R$ is the  matrix for \emph{realization constraints}  i.e., columns correspond to elements of $S_1$,  and rows are of size 
$|H_1|+1$.  Intuitively, the first row of $R$ and $r$  implements $\beta_i(\emptyset)=1$, and the remaining rows implement Equation~\ref{eq:realseq} in the form of for $-\beta_1(\sigma) + \sum_{\move \in \movez_h}\beta_1(\sigma\move)$ = 0 for every $h \in H_1$ and 0 is the zero vector. And optimal mixed-strategy is a strategy that is a solution for the LP given in Equation~\ref{eq:LP}. Realize that the size of LP is linear in the size of the game tree. However, game tree grows exponentially for a given ID, realising every valuation of its conditional dependency tables. Hence, hardness remains. In parallel, recall that  mixed-strategies also include pure strategies. Hence, PSpace-completeness remains. 
Yet one can easily modify the LP given above to fully-mixed strategies by setting $\mu_1 > 0$ i.e., requiring every component of the strategy to be greater than zero.
To apply these results to \IDEL, we simply modify the linear program to consider the evidence of the context
that is given by the observations of the results. Hence, all problems are still solvable in polynomial space.


\section{Related Work}
In addition to the probabilistic logics mentioned in the introduction, some earlier works \cite{acar2015towards,acar2017multi} 
employed (probabilistic) DLs in a decision-theoretic setting. However, neither 
addressed observations, nor contextual reasoning. Hence, they stay completely orthogonal to our work.

Earlier work \cite{koller2003multi,zhou2013game} has used IDs in a game-theoretic setting, yet in a different direction: to 
represent sequential games with more than $n\geq 2$ players compactly and to solve them.
We borrow (in Section 4) the notion of \emph{game-tree} from game theory to compute arbitrary strategies in IDs. There we 
simulate the ID as a 2-player game (against nature) in a game-tree, which allows us to employ linear programming based solution.  
These works also do not consider contextual reasoning (since this is not their motivation).  
To our knowledge, the closest work is  \cite{CePe14} which is of no surprise, since we propose its decision-theoretic extension.

\section{Conclusions}

We introduced \IDEL, a new extension of the light-weight DL \EL capable of modeling and dealing with decision 
situations under uncertainty. This is achieved by integrating an influence diagram to represent the uncertainty, potential decisions, 
and the overall costs of a choice into the knowledge base. The ontological (\EL) portion and the influence representation are 
combined through the use of contexts, which express the situations in which knowledge is required to hold. From an abstract
point of view, we build a collection of ontologies, which are necessarily true only in specific contexts; but, in line with the open-world 
assumption, could still be verified in other situations. These ontologies contain only certain knowledge (i.e., there is no
mention of uncertainty within the ontological knowledge), but the specific context under consideration is uncertain.

Extending the basic idea of the probabilistic DL \BEL, our framework allows for an agent to influence its context through choices in
specific nodes of the network, trying to minimise the overall expected cost over the network. 
Intuitively, this means minimising the probability of large costs, and maximising the probability of low costs. Obviously,
the framework remains uncertain, and there is no absolute guarantee that the observation of the environment do yield the 
lowest possible cost. But the agent can only influence its own choices, not those of the environment.

For this paper, we studied the basic reasoning problems in this logic, and gave tight complexity bounds for all of them. Interestingly, 
the decision problem associated with finding a dominating optimal strategy, in which the agent should find the best strategy 
conditioned on an ontological observation, remains {\sc PSpace}-complete, just as in making inferences over an ID. A practical 
algorithm for solving this problem---and its effective implementation---is left for future work. As future work we will also consider 
other decision-based reasoning tasks, and complexity classes. Notably, we will study whether optimal strategies or costs can be 
approximated efficiently, and whether reasoning becomes tractable over some given parameters. We note that this is
still an open problem even for the special case of \BEL. For example, inferences on a Bayesian network are tractable over
a bounded tree-width, but this property is lost in the currently known algorithms for reasoning in \BEL \cite{CePe-StarAI14}.

Another task to consider is that of building strategies iteratively, as a response to the environment; this is justified by the 
no-forgetting assumption of IDs, and allows an agent to react to newer observations, rather than designing an overall
strategy from the beginning. Some of the complexity issues highlighted in this paper can be alleviated in this way.
Another interesting issue to resolve is how to dislodge the strategies from the underlying ID, and allow the agent to select 
consequences (rather than direct contexts) instead.

To conclude, we note that the choice of \EL as a logical formalism is motivated by its polynomial-time reasoning problems, which
allow us to understand complexity issues better. Likewise, considering TBoxes exclusively, without the addition of ABoxes was
a design choice to simplify the introduction of the formalism. However, our framework can be combined with other (potentially more 
expressive) logics, akin to what was done for Bayesian DLs \cite{CePe17,BoMP19}. Building those extensions introduces further 
problems (e.g., consistency) that would need to be studied in detail as well.

\bibliographystyle{kr}
\bibliography{aaai}

\end{document}